\documentclass[reqno,12pt]{amsart}
\usepackage{amsfonts,amsmath,amssymb,fullpage,cite}
\usepackage{amsaddr}

\usepackage[ansinew]{inputenc}
\usepackage[english]{babel}

\newcommand{\be}{\begin{equation}}
\newcommand{\ee}{\end{equation}}
\newcommand{\bear}{\begin{array}}
\newcommand{\eear}{\end{array}}
\newcommand{\eps}{\varepsilon}

\newcommand{\fif}{\varphi}
\newcommand{\phik}{\sqrt{x^2+(k\pi)^2}}

\newcommand{\dif}{\,\mathrm{d}}
\renewcommand{\i}{\mathrm{i}}
\newcommand{\e}{\mathrm{e}}

\begin{document}

\makeatletter
\title{Infinite series representations for Bessel functions of the first kind of integer order}
\author{A. Andrusyk}
\address{Institute for Condensed Matter Physics, 1 Svientsitsky Street, 79011 Lviv, Ukraine}
\email{aaz@icmp.lviv.ua}

\date{\today}
\keywords{Bessel functions}

\begin{abstract}
We have discovered three non-power infinite series representations for Bessel functions of the first kind of integer orders and real arguments. These series contain only elementary functions and are remarkably simple. Each series was derived as a Fourier series of a certain function that contains Bessel function. The series contain parameter $b$ by setting which to specific values one can change specific form of series. Truncated series retain
qualitatively behaviour of Bessel functions at large $x$: they have sine-like shape with decreasing amplitude. Derived series allow to obtain new series expansions for trigonometric functions.
\end{abstract}

\maketitle

\section*{Introduction}
Bessel functions are widely used within mathematical physics, which explains everlasting interest to these functions both in physics and mathematics. Key properties of Bessel functions were articulated about a hundred years ago. Those results are mainly short, good looking and represent classic results of special function theory. They are collected in numerous treatises (e.g., see~\cite{Watson_1_1995}).

Often works on Bessel functions stem from the needs of physics and are focused on the properties of Bessel functions required to deal with a particular physical problem.
That was the reason to study Bessel functions of large order and argument~\cite{Olver_1_1954,Ratis_1_1993,Paris_1_2004} which arise in astrophysics~\cite{Pierro_1_2001,chistie_1_2008}.

Most works on Bessel functions --- discussing algorithms for Bessel functions calculation~\cite{Coleman_1_1980,Guerrero_1_1988,Ratis_1_1993,Karatsuba_1_1993,Jentschura_1_2012}, identities containing Bessel functions~\cite{Miano_1_1995,Babusci_1_2011,Babusci_2_2011,Dominici_1_2012} and others --- yield cumbersome results. The result of present work, obtained for Bessel functions, is concise and simple.

\section{Infinite series representations for Bessel functions of real order}
We start from three known identities~\cite[2.12.21]{Prudnikov_1_v2_1986}):
\begin{subequations}
\label{eq:integrals0}
\begin{align}
\label{eq:integralA}
&\int_{0}^1
x^{\nu+1}J_{\nu}(bx)\sin{\big(y\sqrt{1-x^2}\big)}\dif x=\sqrt{\frac{\pi}{2}}yb^\nu\big(\sqrt{b^2+y^2}\big)^{-\nu-\frac{3}{2}}J_{\nu+\frac{3}{2}}\big(\sqrt{b^2+y^2}\big),\\
\label{eq:integralB}
&\int_{0}^1
\frac{J_{\nu}(bx)}{\sqrt{1-x^2}}\cos\big(y\sqrt{1-x^2}\big)\dif x=
\frac{\pi}{2}J_{{\nu}/{2}}\bigg(\frac{\sqrt{b^2+y^2}-y}{2}\bigg)J_{{\nu}/{2}}\bigg(\frac{\sqrt{b^2+y^2}+y}{2}\bigg),\\
\label{eq:integralC}
&\int_{0}^1
\frac{x^{\nu+1}J_{\nu}(bx)}{\sqrt{1-x^2}}
\cos\big(y\sqrt{1-x^2}\big)\dif x=
\sqrt{\frac{\pi}{2}}b^{\nu}\big(\sqrt{b^2+y^2}\big)^{-\nu-\frac12}J_{\nu+\frac12}\big(\sqrt{b^2+y^2}\big),
\end{align}
\end{subequations}
where $b>0$, $y$ is any real number and $\nu>-1$. Left sides of Eqs.~(\ref{eq:integrals0}) can be transformed into another form by introducing a new variable $t=\sqrt{1-x^2}$:
\begin{subequations}
\label{eq:integrals1}
\begin{align}
\label{eq:integral1A}
&\int_{-1}^1
t\big(\sqrt{1-t^2}\big)^\nu
J_{\nu}\big(b\sqrt{1-t^2}\big)\sin{(yt)}\dif t=
\sqrt{2\pi}yb^{\nu}\big(\sqrt{b^2+y^2}\big)^{-\nu-\frac{3}{2}}J_{\nu+\frac{3}{2}}\big(\sqrt{b^2+y^2}\big),\\
\label{eq:integral1B}
&\int_{-1}^1
\frac{J_{\nu}\big(b\sqrt{1-t^2}\big)}{\sqrt{1-t^2}}\cos(yt)\dif t=
{\pi}J_{{\nu}/{2}}\bigg(\frac{\sqrt{b^2+y^2}-y}{2}\bigg)J_{{\nu}/{2}}\bigg(\frac{\sqrt{b^2+y^2}+y}{2}\bigg),\\
\label{eq:integral1C}
&\int_{-1}^1
{J_{\nu}(b\sqrt{1-t^2})}
\big({\sqrt{1-t^2}}\big)^{\nu}
\cos(yt)\dif t=
\sqrt{2\pi}b^{\nu}\big(\sqrt{b^2+y^2}\big)^{-\nu-\frac12}J_{\nu+\frac12}\big(\sqrt{b^2+y^2}\big).
\end{align}
\end{subequations}
Here we extended interval of integration over $t$ from $[0,1]$ to $[-1,1]$.

Eq.~(\ref{eq:integrals1}) can be presented in general form
\begin{equation}
\label{eq:integrals2}
\int_{-1}^1 F_{\nu}^{(\alpha)}(b,t)\e^{\i yt}\dif t=f_{\nu}^{(\alpha)}(b,y),\qquad\alpha=A,B,C,
\end{equation}
where
\begin{subequations}
\label{eq:abc}
\begin{align}
\label{eq:a}
\begin{split}
&F_{\nu}^{(A)}(b,t)=-\i t\big(\sqrt{1-t^2}\big)^\nu J_{\nu}\big(b\sqrt{1-t^2}\big),\\
&f_{\nu}^{(A)}(b,y)=\sqrt{2\pi}yb^{\nu}\big(\sqrt{b^2+y^2}\big)^{-\nu-\frac{3}{2}}J_{\nu+\frac{3}{2}}\big(\sqrt{b^2+y^2}\big),
\end{split}
&
\\
\label{eq:b}
\begin{split}
&F_{\nu}^{(B)}(b,t)=\frac{J_{\nu}\big(b\sqrt{1-t^2}\big)}{\sqrt{1-t^2}},\\
&f_{\nu}^{(B)}(b,y)={\pi}J_{{\nu}/{2}}\bigg(\frac{\sqrt{b^2+y^2}-y}{2}\bigg)J_{{\nu}/{2}}\bigg(\frac{\sqrt{b^2+y^2}+y}{2}\bigg),
\end{split}
&
\\
\label{eq:c}
\begin{split}
&F_{\nu}^{(C)}(b,t)={J_{\nu}\big(b\sqrt{1-t^2}\big)}\big({\sqrt{1-t^2}}\big)^{\nu},\\
&f_{\nu}^{(C)}(b,y)=\sqrt{2\pi}b^{\nu}\big(\sqrt{b^2+y^2}\big)^{-\nu-\frac12}J_{\nu+\frac12}\big(\sqrt{b^2+y^2}\big).
\end{split}
&
\end{align}
\end{subequations}

If $y=k\pi$, one can consider integrals on the left side of Eq.~(\ref{eq:integrals2}) to be calculations of the Fourier series coefficients of function $F_{\nu}^{(\alpha)}(b,t)$
defined on interval $t\in(-1,1)$ and extended to a periodic function on $\mathbb{R}$. Therefore one can present $F_{\nu}^{(\alpha)}(b,t)$ in form of the Fourier series
\begin{equation}
\label{eq:series1}
F_{\nu}^{(\alpha)}(b,t)=\frac12\sum\limits_{k=-\infty}^{\infty}f_{\nu}^{(\alpha)}(b,k\pi)\e^{-\i k\pi t},
\begin{array}{ll}
\quad t\in(-1,1)& \mbox{if\ \ } \alpha=A\;\mbox{and}\:-1<\nu\leq0, \\
\quad t\in[-1,1]& \mbox{if\ \ } \alpha=A\;\mbox{and}\:\nu>0,\,\mbox{or}\;\alpha=B,C.
\end{array}
\end{equation}
In case of a singularity at $t=\pm1$ on the left side of Eq.~(\ref{eq:series1}) corresponding series on the right side will also give singularity. In case of the removable singularity at $t=\pm1$ we have to remove singularity by redefining the function at these points and the equation will remain true.
Series in Eq.~(\ref{eq:series1}) gives zero
at points $t=\pm1$ if $\alpha=A$ and $-1<\nu\leq0$. All Fourier series (\ref{eq:series1}) are valid while $F_{\nu}^{(A,B,C)}(b,t)$ are differentiable
with respect to $t$ at $t\in(-1,1)$.

Coming back to variable $x=\sqrt{1-t^2}$ Eq.~(\ref{eq:series1}) is transformed into
\begin{equation}
\label{eq:series2}
\begin{split}
&F_{\nu}^{(\alpha)}\big(b,\sqrt{1-x^2}\big)=\frac12\sum\limits_{k=-\infty}^{\infty}
f_{\nu}^{(\alpha)}(b,k\pi)\e^{-\i k\pi\sqrt{1-x^2}},
\quad
\mbox{where}\\
&\begin{array}{ll}
\quad x\in(0,1]& \mbox{if\ \ } \alpha=A\;\;\mbox{and}\:\:-1<\nu\leq0, \\
\quad x\in[0,1]& \mbox{if\ \ } \alpha=A\;\;\mbox{and}\:\:\nu>0,\;\mbox{or}\;\;\alpha=B,C.
\end{array}
\end{split}
\end{equation}
Exchanging $b$ and $x$ we get the following series expansions at $x>0$
\begin{equation}
\label{eq:series3}
\begin{split}
&F_{\nu}^{(\alpha)}\big(x,\sqrt{1-b^2}\big)=\frac12\sum\limits_{k=-\infty}^{\infty}
f_{\nu}^{(\alpha)}(x,k\pi)\e^{-\i k\pi\sqrt{1-b^2}},
\quad
\mbox{where}\\
&\begin{array}{ll}
\quad b\in(0,1]& \mbox{if\ \ } \alpha=A\;\;\mbox{and}\:\:-1<\nu\leq0, \\
\quad b\in[0,1]& \mbox{if\ \ } \alpha=A\;\;\mbox{and}\:\:\nu>0,\;\mbox{or}\;\;\alpha=B,C,
\end{array}
\end{split}
\end{equation}
which in every particular case of $\alpha$ has the form
\begin{equation}
\label{eq:expansionA}
b^{\nu}\sqrt{1-b^2}J_{\nu}(bx)=\sum\limits_{k=1}^{\infty}f_{\nu}^{(A)}(x,k\pi)\sin\big(k\pi\sqrt{1-b^2}\big),
\begin{array}{ll}
\quad b\in(0,1]& \mbox{if\ \ }-1<\nu\leq0, \\
\quad b\in[0,1]& \mbox{if\ \ }\nu>0,
\end{array}
\end{equation}
which we will call (A)-case,
\begin{equation}
\label{eq:expansionB}
\frac{J_{\nu}(bx)}{b}=\sum\limits_{k=0}^{\infty}\eps_k
f_{\nu}^{(B)}(x,k\pi)\cos\big(k\pi\sqrt{1-b^2}\big),\quad b\in[0,1],
\end{equation}
which we will call (B)-case, and
\begin{equation}
\label{eq:expansionC}
b^{\nu}J_{\nu}(bx)=\sum\limits_{k=0}^{\infty}\eps_k
f_{\nu}^{(C)}(x,k\pi)\cos\big(k\pi\sqrt{1-b^2}\big),\quad b\in[0,1],
\end{equation}
which we will call (C)-case, where
\begin{equation}
\label{eq:eps}
\eps_k=
\begin{cases}
\frac12, & \mbox{if } k=0 \\
1, & \mbox{if } k>0.
\end{cases}
\end{equation}

Eqs.~(\ref{eq:expansionA}), (\ref{eq:expansionB}), (\ref{eq:expansionC}) are valid at $x>0$ and $\nu>-1$.

\section{Infinite series representations for Bessel functions of integer order}
Derived series representations merit detailed consideration at integer $\nu$ as in this case
they get series representations through the elementary functions. Series representations for Bessel function of integer order are following:

(A)-case
\begin{equation}
\label{eq:expansionA_n}
b^{n}J_{n}(bx)=\sum\limits_{k=1}^{\infty}(k\pi)g_k^{(A)}(b)f_{n}^{(A)}(x,k\pi),\quad b\in(0,1),\quad n\geq0,\quad x\geq0,
\end{equation}

(B)-case\footnote{Here we will not receive series representation for $J_0(x)$ through elementary functions (see Eqs.~(\ref{eq:expansionB}) and~(\ref{eq:b})). Therefore we consider only case of $n\geq1$.}\footnote{Left side of this equation should be taken as $\lim_{b\to0}\frac{J_n(bx)}{b}$ at $b=0$.}
\begin{equation}
\label{eq:expansionB_n}
\frac{J_{n}(bx)}{b}=\sum\limits_{k=0}^{\infty}\eps_k
g_k^{(B,C)}(b)f_{n}^{(B)}(x,k\pi),\quad
b\in[0,1],\quad n\geq1,\quad x\geq0,
\end{equation}

(C)-case
\begin{equation}
\label{eq:expansionC_n}
b^{n}J_{n}(bx)=\sum\limits_{k=0}^{\infty}\eps_k
g_k^{(B,C)}(b)f_{n}^{(C)}(x,k\pi),\:
\begin{array}{ll}
\quad b\in[0,1]& \mbox{if\ \ }n=0, \\
\quad b\in(0,1]& \mbox{if\ \ }n\geq1,
\end{array},\quad x\geq0,
\end{equation}
where
\begin{equation}
\label{eq:gka_gkbc}
g_k^{(A)}(b)=\frac{\sin\big(k\pi\sqrt{1-b^2}\big)}{k\pi\sqrt{1-b^2}},
\qquad
\mbox{and}
\qquad
g_k^{(B,C)}(b)=\cos\big(k\pi\sqrt{1-b^2}\big).
\end{equation}
We imply that $f_{n}^{(\alpha)}(0,k\pi)=\lim_{x\to0}f_{n}^{(\alpha)}(x,k\pi)$ ($\alpha=A,B,C$)
if singularity at $x=0$ is present.

However, we need to perform some transformations in (B)-case and at $n=2m$ in order to get elementary functions on the right hand
of Eq.~(\ref{eq:expansionB}). Application of recurrence relations for the Bessel functions \cite{Watson_1_1995}
\begin{equation}
\label{eq:recurrence}
\frac{4m}{b^2x}J_{2m}(bx)=\frac{J_{2m-1}(bx)}{b}+\frac{J_{2m+1}(bx)}{b},\quad
\;m\geq1
\end{equation}
to equation (\ref{eq:expansionB}) results in representation\footnote{We should note that there are
numerous ways to derive Bessel functions of even order through Bessel functions of odd order. For instance, recurrent relations \cite{Watson_1_1995} give another two equations:
\[
\begin{split}
J_{2m}(bx)&=(2m+1)J_{2m+1}(bx)+\frac{1}{b}\frac{\dif J_{2m+1}(bx)}{\dif x},\\
J_{2m}(bx)&=(2m-1)J_{2m-1}(bx)-\frac{1}{b}\frac{\dif J_{2m-1}(bx)}{\dif x}.
\end{split}
\]}
\begin{subequations}
\label{eq:expansionB_even}
\begin{align}
\label{eq:exp_JB_even}
&\frac{4m}{b^2}J_{2m}(bx)=\sum\limits_{k=0}^{\infty}\eps_k
g_k^{(B,C)}(b)f_{2m}^{(B)}(x,k\pi),\quad b\in(0,1],\quad x\geq0,\quad m\geq1\\
\label{eq:exp_fB_even}
\begin{split}
f_{2m}^{(B)}(x,y)={}x\cdot\pi
\bigg[
J_{(2m-1)/2}\bigg(\frac{\sqrt{x^2+y^2}-y}{2}\bigg)J_{{(2m-1)}/{2}}\bigg(\frac{\sqrt{x^2+y^2}+y}{2}\bigg)\\
+J_{(2m+1)/2}\bigg(\frac{\sqrt{x^2+y^2}-y}{2}\bigg)J_{{(2m+1)}/{2}}\bigg(\frac{\sqrt{x^2+y^2}+y}{2}\bigg)
\bigg].
\end{split}
\end{align}
\end{subequations}

We can extend derived expansions to the region $x<0$ according to the parity of the Bessel functions of integer order: $J_{n}(-x)=(-1)^nJ_{n}(x)$.

With notation
\begin{equation}
\label{eq:phi}
\varphi_k=\phik
\end{equation}
the series expansions for the Bessel functions of order $0$, $1$, $2$ 
are presented below for every particular case.

(A)-case:
\begin{subequations}
\label{eq:particular_b_A}
\begin{align}
\label{eq:Bess_b_A_0}
&J_0(bx)=2\sum\limits_{k=1}^{\infty}(k\pi)^2g_k^{(A)}(b)\frac{\sin{\varphi_k}-{\varphi_k}\cos{\varphi_k}}{{\varphi_k}^3},\\
\label{eq:Bess_b_A_1}
&J_1(bx)=2\frac{x}{b}\sum\limits_{k=1}^{\infty}(k\pi)^2g_k^{(A)}(b)\frac{(3-{{\varphi_k}^2})\sin{\varphi_k}-3{\varphi_k}\cos{\varphi_k}}{{\varphi_k}^5},\\
\label{eq:Bess_b_A_2}
&J_2(bx)=2\frac{x^2}{b^2}\sum\limits_{k=1}^{\infty}(k\pi)^2g_k^{(A)}(b)\frac{{\varphi_k}({\varphi_k}^2-15)\cos{\varphi_k}+3(5-2{\varphi_k}^2)\sin{\varphi_k}}{{\varphi_k}^7},
\end{align}
\end{subequations}
where $b\in(0,1)$.

(B)-case:
\begin{subequations}
\label{eq:particular b_B}
\begin{align}
\label{eq:Bess b_B_1}
&J_1(bx)=\frac{2b}{x}\sum\limits_{k=0}^{\infty}\eps_k{g_k^{(B,C)}(b)}\big[(-1)^k-\cos{\varphi_k}\big],\\
\label{eq:Bess_b_B_2}
&J_2(bx)=\frac{b^2}{x^2}\sum\limits_{k=0}^{\infty}\eps_k{g_k^{(B,C)}(b)}\big[
(-1)^k(2+x^2)-2(\varphi_k\sin{\varphi_k}+\cos{\varphi_k})
\big],
\end{align}
\end{subequations}
where $b\in[0,1]$.

(C)-case:
\begin{subequations}
\label{eq:particular_b_C}
\begin{align}
\label{eq:Bess_b_C_0}
&J_0(bx)=2\sum\limits_{k=0}^{\infty}\eps_k{g_k^{(B,C)}(b)}\frac{\sin{\varphi_k}}{\varphi_k},\\
\label{eq:Bess_b_C_1}
&J_1(bx)=2\frac{x}{b}\sum\limits_{k=0}^{\infty}\eps_k{g_k^{(B,C)}(b)}\frac{\sin{\varphi_k}-{\varphi_k}\cos{\varphi_k}}{{\varphi_k}^3},\\
\label{eq:Bess_b_C_2}
&J_2(bx)=2\frac{x^2}{b^2}\sum\limits_{k=0}^{\infty}\eps_k{g_k^{(B,C)}(b)}\frac{(3-{{\varphi_k}^2})\sin{\varphi_k}-3{\varphi_k}\cos{\varphi_k}}{{\varphi_k}^5},
\end{align}
\end{subequations}
where $b\in[0,1]$ for $J_0(bx)$, and $b\in(0,1]$ for $J_n(bx)$, $n\geq1$.

One can derive following properties of the $k$-th term in series (\ref{eq:expansionA_n}), (\ref{eq:expansionB_n}), (\ref{eq:expansionC_n}) at $k\to\infty$:
\begin{subequations}
\label{eq:decrease_A}
\begin{align}
\label{eq:decrease0_A}
&\qquad\Big|(k\pi)g_{k}^{(A)}(b)f_{n}^{(A)}(x,k\pi)\Big|\leq\Big|(k\pi)f_{n}^{(A)}(x,k\pi)\Big|,\quad
n,m=\overline{0,\infty},\\
\label{eq:decrease1_A}
&(k\pi)f_{2m}^{(A)}(x,k\pi)\sim2(-1)^{k+m+1}\left(\frac{x}{k\pi}\right)^{2m},\\
\label{eq:decrease2_A}
&(k\pi)f_{2m+1}^{(A)}(x,k\pi)\sim2\frac{(-1)^{k+m+1}}{k\pi}\left(\frac{x}{k\pi}\right)^{2m+1} \big[x^2+(2m+2)(2m+3)\big];
\end{align}
\end{subequations}
\begin{subequations}
\label{eq:decrease_B}
\begin{align}
\label{eq:decrease0_B}
&\qquad\Big|g_{k}^{(B,C)}(b)f_{n}^{(B)}(x,k\pi)\Big|\leq\Big|f_{n}^{(B)}(x,k\pi)\Big|,\quad
n=\overline{1,\infty},\quad m=\overline{0,\infty},\\
\label{eq:decrease1_B}
&f_{4m+1}^{(B)}(x,k\pi)\sim(-1)^{k+m}\frac{x^{2m-1}}{2^{2m+1}}\left(\frac{x}{k\pi}\right)^{2m+2}\frac{(2m+1)!}{(2(2m+1))!}\big[x^2+4m(2m+1)\big],\quad\\
\label{eq:decrease2_B}
&f_{4m+2}^{(B)}(x,k\pi)\sim(-1)^{k+m}\frac{x^{2m}}{2^{2m}}\left(\frac{x}{k\pi}\right)^{2m+2}\frac{(2m+1)(2m+1)!}{(4m+3)(2(2m+1))!}\big[x^2+2m(4m+3)\big],\\
\label{eq:decrease3_B}
&f_{4m+3}^{(B)}(x,k\pi)\sim{(-1)^{k+m+1}}\frac{x^{2m+1}}{2^{2m}}\left(\frac{x}{k\pi}\right)^{2m+2}\frac{(2m+2)!}{(2(2m+2))!},\\
\label{eq:decrease4_B}
&f_{4m+4}^{(B)}(x,k\pi)\sim{(-1)^{k+m+1}}\frac{x^{2m+2}}{2^{2m}}\left(\frac{x}{k\pi}\right)^{2m+2}\frac{(2m+2)!}{(2(2m+2))!};
\end{align}
\end{subequations}
\begin{subequations}
\label{eq:decrease_C}
\begin{align}
\label{eq:decrease0_C}
&\qquad\Big|g_{k}^{(B,C)}(b)f_{n}^{(C)}(x,k\pi)\Big|\leq\Big|f_{n}^{(C)}(x,k\pi)\Big|,\quad
n,m=\overline{0,\infty},\\
\label{eq:decrease1_C}
&f_{2m}^{(C)}(x,k\pi)\sim(-1)^{k+m}\frac{1}{(k\pi)^2}\left(\frac{x}{k\pi}\right)^{2m} \big[x^2+2m(2m+1)\big],\\
\label{eq:decrease2_C}
&f_{2m+1}^{(C)}(x,k\pi)\sim2(-1)^{k+m+1}\frac1{k\pi}\left(\frac{x}{k\pi}\right)^{2m+1}.
\end{align}
\end{subequations}
Notation $a_k\sim b_k$ means that $\lim_{k\to\infty}\frac{a_k}{b_k}=1$. Eqs.~(\ref{eq:decrease_A}),~(\ref{eq:decrease_B}),~(\ref{eq:decrease_C}) allow to conclude absolute convergence and uniform convergence with respect to parameter $b$ at $b\in(0,1)$ for series (\ref{eq:expansionA_n}), (\ref{eq:expansionB_n}), and (\ref{eq:expansionB_n}) except for series (\ref{eq:expansionA_n}) at $n=0$ (that is (\ref{eq:Bess_b_A_0}))
which is conditionally convergent and is not uniformly convergent with respect to $b$ at $b\in(0,1)$.

Due to the uniform convergence of series~(\ref{eq:expansionA_n}) with respect to $b\in(0,1)$ at $n\geq1$ we can take a termwise limit $b\to1$ and hence put $g_k^{(A)}(b)=1$ at $b\to1$ in (A)-case:

\begin{equation}
\label{eq:expansionA2_n}
J_{n}(x)=\sum\limits_{k=1}^{\infty}(k\pi)f_{n}^{(A)}(x,k\pi),\quad n\geq1,\quad x\geq0.
\end{equation}
Series (\ref{eq:expansionA2_n}) at $n=0$ diverges.

The series representations for the Bessel functions of order $0$, $1$, $2$ 
at $b=1$ are presented below.
(A)-case:
\begin{subequations}
\label{eq:particular_A}
\begin{align}
\label{eq:Bess_A_1}
&J_1(x)=2x\sum\limits_{k=1}^{\infty}(k\pi)^2\frac{(3-{{\varphi_k}^2})\sin{\varphi_k}-3{\varphi_k}\cos{\varphi_k}}{{\varphi_k}^5},\\
\label{eq:Bess_A_2}
&J_2(x)=2x^2\sum\limits_{k=1}^{\infty}(k\pi)^2\frac{{\varphi_k}({\varphi_k}^2-15)\cos{\varphi_k}+3(5-2{\varphi_k}^2)\sin{\varphi_k}}{{\varphi_k}^7},
\end{align}
\end{subequations}
(B)-case:
\begin{subequations}
\label{eq:particular_B}
\begin{align}
\label{eq:Bess_B_1}
&J_1(x)=\frac{2}{x}\sum\limits_{k=0}^{\infty}\eps_k\big[(-1)^k-\cos{\varphi_k}\big],\\
\label{eq:Bess_B_2}
&J_2(x)=\frac{1}{x^2}\sum\limits_{k=0}^{\infty}\eps_k\big[
(-1)^k(2+x^2)-2(\varphi_k\sin{\varphi_k}+\cos{\varphi_k})
\big],
\end{align}
\end{subequations}
(C)-case:
\begin{subequations}
\label{eq:particular_C}
\begin{align}
\label{eq:Bess_C_0}
&J_0(x)=2\sum\limits_{k=0}^{\infty}\eps_k\frac{\sin{\varphi_k}}{\varphi_k},\\
\label{eq:Bess_C_1}
&J_1(x)=2x\sum\limits_{k=0}^{\infty}\eps_k\frac{\sin{\varphi_k}-{\varphi_k}\cos{\varphi_k}}{{\varphi_k}^3},\\
\label{eq:Bess_C_2}
&J_2(x)=2x^2\sum\limits_{k=0}^{\infty}\eps_k\frac{(3-{{\varphi_k}^2})\sin{\varphi_k}-3{\varphi_k}\cos{\varphi_k}}{{\varphi_k}^5}.
\end{align}
\end{subequations}

According to some simple considerations, we conclude that series (\ref{eq:expansionA_n}), (\ref{eq:expansionB_n}), (\ref{eq:expansionC_n}) at all permitted values of parameter $b$, and series (\ref{eq:expansionA2_n}) converge uniformly with respect to $x\in[0,A]$ where $A$ is arbitrary large real number.

One can derive another series representations for Bessel functions by taking $b\neq1$ and scaling variable $x$ by $1/b$. So (A)-case gives the following series representation for $J_0(x)$ at $b=\sqrt{3}/2$:
\begin{equation}
\label{eq:J0_A_case}
\begin{split}
&J_0(x)=4\sum\limits_{n=1}^{\infty}(-1)^n\big[(2n-1)\pi\big]\frac{\psi_n\cos\psi_n-\sin\psi_n}{{\psi_n}^3},\quad\mbox{where}\\
&\psi_n=\sqrt{\frac{4x^2}{3}+\big[(2n-1)\pi\big]^2}.
\end{split}
\end{equation}

\section{Infinite series representations for sine and cosine functions}
Derived series expansions allow to present numerous similar expansions for sine and cosine functions. We will present some of them.

Lets consider Eq.~(\ref{eq:Bess b_B_1}) which we rewrite in the following way:
\begin{equation}
\label{eq:series}
x\frac{J_1(bx)}{b}=2\sum\limits_{k=0}^{\infty}\eps_k{g_k^{(B,C)}(b)}\big[(-1)^k-\cos{\varphi_k}\big],
\end{equation}
where $\varphi_k$ is presented in Eq.~(\ref{eq:phi}).
Series (\ref{eq:series}) is uniformly convergent with respect to $b\in(0,1)$ therefore taking a limit $b\to0$ in Eq.~(\ref{eq:series}) gives us the following result:
\begin{equation}
\label{eq:cos_1}
\cos{x}-1+\frac{x^2}{2}=2\sum\limits_{k=1}^{\infty}
\Big[
1-(-1)^k\cos{\fif_k}
\Big].
\end{equation}

Taking $b=0$ in Eq.~(\ref{eq:Bess_b_C_0}) we derive series for sine function:
\begin{equation}
\label{eq:sin_1}
1-\frac{\sin{x}}{x}=2\sum\limits_{k=1}^{\infty}
\bigg[
(-1)^{k}\frac{\sin{\fif_k}}{\fif_k}
\bigg].
\end{equation}
It can be also derived by differentiating Eq.~(\ref{eq:cos_1}) with
respect to $x$.

Subtracting Eq.~(\ref{eq:Bess_B_1}) and Eq.~(\ref{eq:Bess_C_1}) we get another series for sine function:
\begin{equation}
\label{eq:sin_2}
1-\frac{\sin{x}}{x}=-2\sum\limits_{k=1}^{\infty}
\bigg[
(-1)^k-\frac{(k\pi)^2}{{\fif_k}^2}\cos{\fif_k}-\frac{x^2}{{\fif_k}^2}\frac{\sin\fif_k}{\fif_k}
\bigg]
\end{equation}
which converges faster than the previous one.

Equations (\ref{eq:cos_1}), (\ref{eq:sin_1}), (\ref{eq:sin_2}) can be differentiated any times that will give plenty of other series representations for sine and cosine functions.

\section*{Conclusion}
To sum up we outline key properties of the series representations (\ref{eq:expansionA_n}), (\ref{eq:expansionB_n}), (\ref{eq:expansionC_n}), and (\ref{eq:expansionA2_n}) for Bessel functions of the first kind of integer order:
\begin{itemize}
\item series (\ref{eq:expansionA_n}) at $n\geq1$, (\ref{eq:expansionB_n}), (\ref{eq:expansionC_n}) are uniformly convergent with respect to parameter $b\in(0,1)$ at all $x$;
\item series (\ref{eq:expansionA_n}), (\ref{eq:expansionB_n}), (\ref{eq:expansionC_n}) at all permitted values of $b$, and series (\ref{eq:expansionA2_n}) are uniformly convergent with respect to variable $x\in[0,A]$, where $A$ is arbitrary large real value;
\item series (\ref{eq:expansionA_n}) at $n\geq1$, (\ref{eq:expansionB_n}), (\ref{eq:expansionC_n}) at all permitted values of $b$, and series (\ref{eq:expansionA2_n}) are absolutely convergent; series (\ref{eq:expansionA_n}) at $n=0$ (that is (\ref{eq:Bess_b_A_0})) is conditionally convergent.
\end{itemize}

As series (\ref{eq:expansionB_n}), (\ref{eq:expansionC_n}) at $b=1$ and (\ref{eq:expansionA2_n}) become alternating and $k$-th terms approach zero monotonically at sufficient large $k$ and $x\in[0,A]$ (see Eqs.~(\ref{eq:decrease_A}), (\ref{eq:decrease_B}), (\ref{eq:decrease_C})), the
bonds for the error terms of infinite sums approximations by truncated series to $K$-term are absolute value of ($K+1$)-th term. As one can see, ($K+1$)-th terms are monotonically decreasing functions of $x$ in all three cases. Besides, truncated series (\ref{eq:expansionA_n}), (\ref{eq:expansionB_n}), (\ref{eq:expansionC_n}), (\ref{eq:expansionA2_n}) have oscillating behaviour (sine-like shape) with decreasing amplitude at large $x$ that is they behave exactly like Bessel functions. However, truncated series fail to provide correct asymptotic behaviour of Bessel function at $x\to\infty$~\cite{Watson_1_1995}:
\[
J_n(x)\sim\sqrt{\frac{2}{\pi x}}\cos\Big(x-\frac{n\pi}{2}-\frac{\pi}{4}\Big).
\]

\end{document}